# A 16-pixel NbN nanowire single photon detector coupled with 300 μm fiber

Qi Chen*, Biao Zhang*, Labao Zhang**, Rui Ge, Ruiying Xu, Yang Wu, Xuecou Tu, Xiaoqing Jia, Lin Kang, Jian Chen, Peiheng Wu

Nanjing University, Hankou Road 22, Nanjing 210093, China

*These authors contributed equally to this work.

**Lzhang@nju.edu.cn

**Abstract**

Niobium Nitride (NbN) nanowire is the most popular detection material of superconducting nanowire single photon detectors (SNSPDs) for high repetition rate and high efficiency. However, it has been assumed to be difficult for fabricating SNSPDs with arrays over large area, which are critical components of quantum imaging, linear optical quantum computing, satellite laser ranging, and high-speed quantum key distribution. This paper reported a 4×4 pixel NbN SNSPDs array with an equivalent receiving aperture of ϕ300 μm associated with beam compression technology, which is the first NbN SNSPD coupled by ϕ300 μm fiber according to our knowledge. The designed pixel was compact with a filling factor of 98.5%, resulting in a high quantum efficiency of 94.5%, a system efficiency of 46% for photons coupled from ϕ300 μm fiber without optimizing polarization, and a system time resolution of 92 ps. This work demonstrates the feasibility of a large SNSPDs array, and paves the way for developing efficient photon camera with NbN nanowires.

## 1. Introduction

Superconducting nanowire single photon detectors (SNSPDs) introduced many revolutions in detector performance and scientific applications[1]. Since the first SNSPD (composed of a 1 μm long, 200 nm wide Niobium Nitride (NbN) nanowire[2]), SNSPDs reached 93% detection efficiency[3], sub-10ps time jitter[4], less than 0.01 cps dark count rate[5], and were applied in the loop-free test of Bell equation[6,7], lunar-to-earth laser communication[8,9], quantum imaging[1,10], linear optical quantum computing[11], and high-speed quantum key distribution[12].

NbN is typically selected due to low time jitter[13], low dark count rate[5], high repetition rate[14] and high critical temperature[15] compared to amorphous materials such as WSi[16] and MoSi[17]. Recent work[18] pointed out that NbN based-SNSPDs can also perform excellent with high efficiency over 90%. Therefore, NbN is the most promising material in researching SNSPDs with high speed, high stability, and high efficiency. However, NbN in SNSPDs is polycrystalline or single crystal expital depending on the substrates. Thus it is difficult to produce a uniformed film high-quality over large area due to factors such as crystal orientation and lattice defects[19].

In the past decade, scientists paid much attention to increasing the NbN-nanowire pixel number and active detection area by sacrificing the filling factor, while reducing its detection efficiency (usually below 10%). S. Doerner et al.[20] developed a 16-pixel SNSPDs array with a filling factor of only 14%, resulting in a maximum detection efficiency below 10%. Divochiy et al.[21] proposed a parallel nanowire detector with a low filling factor, with a quantum efficiency of only about 2% at 1.3 μm. Jahanmirinejad et al.[22] developed a nanowires array that covered a total active area of only 12 μm × 12 μm. Miki et al.[23] proposed a 64-pixel SNSPD with a filling factor of 40% over 63 μm × 63 μm. Fabrication NbN-nanowire array with high efficiency requires much high chip yield. It was even concluded that NbN is unsuitable to prepare SNSPDs with large active area with NbN[3,16].

In this work, a NbN-based SNSPD array was fabricated that consists of 16 elements that area arranged in a 4×4 array, where every single pixel has a size of 20 μm × 20 μm and the total active area is 80 μm × 80 μm on a silicon wafer with $Si_3N_4$ anti-reflection layer. Wiring is compactly arranged between pixels, which ensuring a filling rate of 98.5%. The coupling beam in which the optical path is filled with multimode fibers and processed beam compression technology has a diameter of up to 300 μm, which reduces the difficulty of optical coupling and photon loss caused by the coupling process; consequently, a system detection efficiency (SDE) of 46% and a quantum efficiency (QE) of 94.5% can be reached.

## 2. Experimental
### Device design

The optimization of the device structure, refers to the design of Min et al.[24]. The traditional high-Q optical cavity structure is adopted for the SNSPD structural design, and FDTD solutions simulation software was used to optimize the device structure parameters. Fig. 1(a) depicts the simulation model. The substrate material is Si. To enhance the light absorption ratio of the nanowire, not only a $Si_3N_4$ anti-reflection film is adopt, but also a thin layer of $Si_3N_4$ with a thickness of 135 nm is added to the model, which is used as a resonator in lower layers. The nanowires are placed above the lower resonant cavity. To further enhance the light absorption ratio of the nanowire, the upper $Si_3N_4$ resonant cavity is added above the nanowires. Furthermore, a 200 nm thick gold mirror is prepared on the upper resonant cavity to completely reflect photons. In the structure of the device, the upper and lower layers of $Si_3N_4$ form a complete optical cavity. In the simulation process, the cavity structure boundary conditions are set as follows: the periodic boundary conditions are set in both x and y directions, while the z direction is set as a perfectly matched layer (PML). The photons enter the device from the back side of the Si substrate, the light is polarized, and the polarization directions are parallel and perpendicular to the nanowires. Combining theory and practice, the nanowire linewidth is 100 nm, the filling factor is 1/3, and the film thickness is 6.5 nm. To facilitate rapid simulation while obtaining high precision simulation results at the same time, further mesh refinement operation on the part of NbN nanowires was conducted. The grid precision in the x, y, and z triaxial directions are set to 0.5 nm. Only x = 0 is simulated with a two-dimensional plane composed of the y-axis. A wavelength λ range from 600 nm to 1600 nm was simulated. The refractive index n of each simulated material varies with wavelength, when at λ = 1064 nm, $nSi = 3.4215$; $nSi_3N_4 = 2.0112$; $nNbN = 4.1141+4.9423i$; $nAu = 0.28+9.03i$.

To maximize the absorption rate of photons by NbN nanowires, it is necessary to optimize the thickness d of the upper resonator via simulation to achieve critical coupling. Changing the thickness of the upper resonator, also changes the center wavelength. After optimization, a relatively suitable thickness of d = 133 nm was obtained. The simulation results are shown in Fig. 1(b). With changing wavelength, the NbN nanowires almost changed the absorption rate of TE and TM waves. The absorption of polarized waves by NbN nanowires ranges between 70% and 80% at 1064 nm, and the device is insensitive to the polarization of light. Fig. 1(c) shows a three-dimensional structure of the SNSPD. The yellow structure represents an Au electrode. In Fig.1(d), the purple part represents the nanowire array, an Au mirror is installed above it, and the 16 nanowire pixels are respectively taken out by 16 Au short pins. The electrode numbers are marked from 1 to 16. To facilitate the layout of the nanowire array and the subsequent device test analysis, 10 long pins are designed. Between 16 Au short pins and two large size pins are designed in the middle of the device to lead out the G levels of all nanowire pixels. At the same time, the device is connected to the grounded outer frame to achieve common ground.

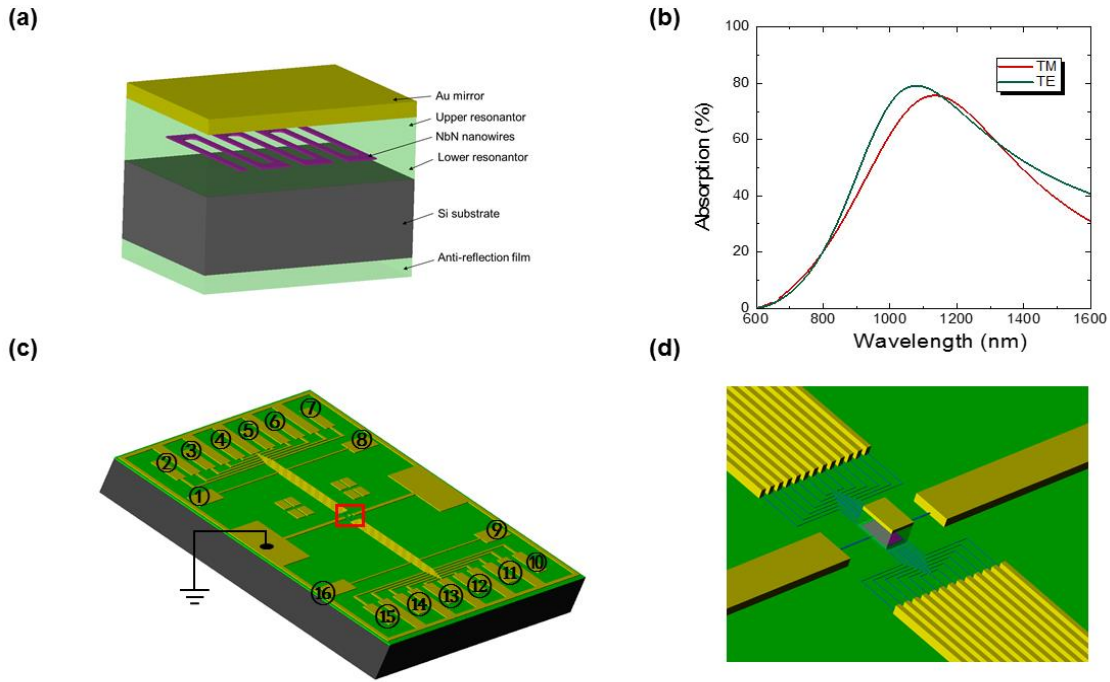

Fig. 1. Device design and simulation diagrams. (a) The simulated structure is shown. (b) Absorption of NbN nanowire vs wavelength from 600 nm to 1600 nm. With changing wavelength, the NbN nanowires almost change the absorption rate of TE and TM waves. (c) Three-dimensional structure of the SNSPD, the red solid line box marks the site for detection area of the SNSPD. (d) Magnification of the nanowire detection area, the purple part represents the nanowire array.

**Optimization of the fabrication process**

In this experiment, a nanowire width of 100 nm is designed with a filling factor of 1/3. The total active area of the superconducting nanowires increases to 80 μm × 80 μm, and the single pixel is 20 μm × 20 μm. Compact wiring between the cells ensures a 98.5% filling rate. For the convenience of test and analysis, each pixel number is defined (such as #1, #2,...) in Fig. 2.

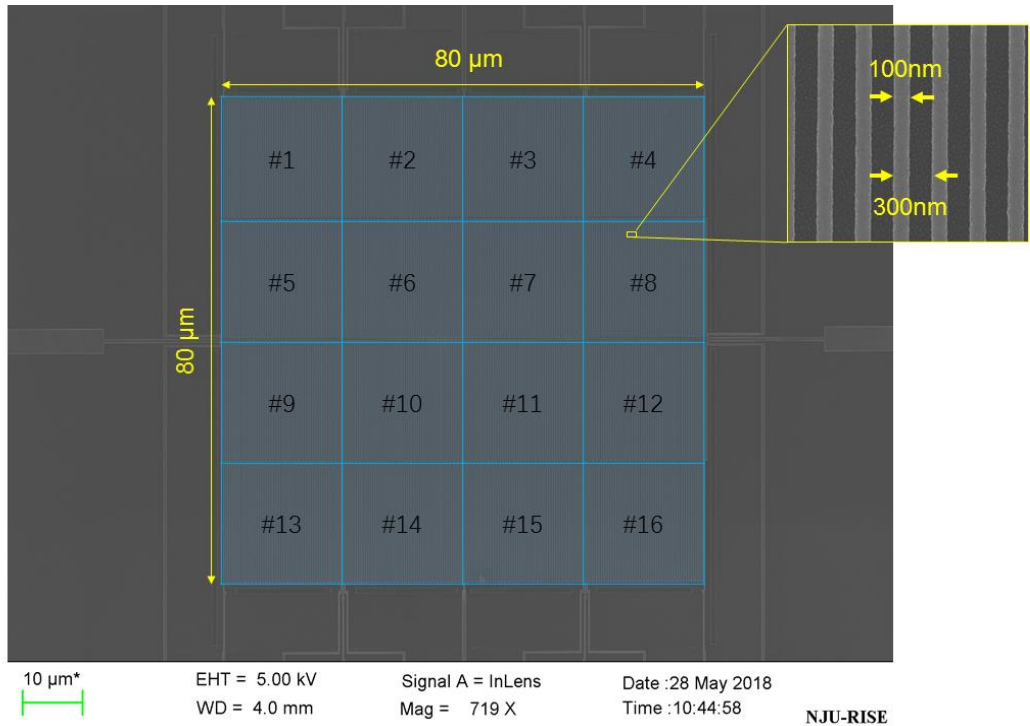

Fig. 2. Nanowire array photographs taken via scanning electron microscope (SEM).

The nanowire preparation process is based on previous work[25–27] and the materials and micro-nanofabrication processing were further optimized. First, by optimizing the PECVD process, the refractive index of $Si_3N_4$ at λ = 1064 nm of 2.0112 was ensured, as well as a surface flatness Root-Mean-Square (RMS) < 0.5 nm. A 135 nm $Si_3N_4$ layer was grown on the front side of the wafer as a lower resonator with a growth temperature of 350 ℃ using a Plasma80Plus device from Oxford Instruments. To reduce the reflectivity of the light incident from the back side of the device, a 135 nm thick $Si_3N_4$ layer was grown on the back side of the wafer. Next, a NbN film was deposited on the lower resonator via a DC magnetron sputtering(model number DE500). The thickness of the NbN film was 6.5 nm, and the superconducting critical temperature of the film was 8.3 K. Then, the Au electrode was prepared via photolithography (the Au electrode can achieve low loss transmission of the RF signal, and can also help adjust the incident beam to the device detection area during the packaging process). A two-layer adhesive lithography process was used. The lower layer of glue is LOR10B and the upper layer is AZ1500. The Au electrode was grown via using a DC magnetron sputtering device(model number DE500). The thickness of the Au electrode is about 200 nm. To enhance the adhesion of the Au electrode to NbN, a layer of Ti was grown with a thickness of about 10 nm between Au and NbN. After the lift-off process, the Au electrode was prepared and the electron beam exposure was started.

The substrate was spin-coated with a 2% concentration of hydrogen silsesquioxane (HSQ), which is an electron beam photoresist[28] manufactured by Dow Corning. The spinner rotation speed was controlled to about 5000 rpms and the spin coating time was 1 min. After the spin coating was completed, the substrate was baked on a drying table at 100 ℃ for 3 min. After baking, the substrate was left to return to room temperature and was then subjected to electron beam lithography. The electron beam lithography system EBPG5200 from Raith was selected. During the electron beam lithography process, the electron beam current was set to 0.2 nA and the acceleration voltage was 100 KV. These values were chosen to ensure that the quality of the nanowires generated by the exposure is excellent while the exposure time is not excessively long. In addition, since the electron proximity effect is particularly severe during large area exposure, a reasonable control of the exposure dose is important. Fig.3 shows the relationship between this exposure dosage and the nanowire linewidth during electron beam lithography, with doses ranging from 1100 μC/cm² to 1800 μC/cm².

For the convenience of analysis, as shown in Fig. 3(a), four pixels (pixel 2, pixel 3, pixel 11, and pixel 10) were selected. Fig. 3(b) shows the different nanowire linewidths and the standard deviations for the four pixels at different exposure dosages. As shown in Fig. 3(b), the nanowires obtained by both pixel 2 and pixel 10, or both pixel 3 and pixel 11 at the same exposure dosage have the same linewidths. This indicates that the electron beam lithography at different positions is consistent without considering the electron beam proximity effect. Moreover, due to the electron beam proximity effect, the nanowire widths of pixel 3 and pixel 11 exceed the edge portions of pixel 2 and pixel 10 to different extents for different exposure dosages. At the same time, when the exposure dosage is too high, the width uniformity of the nanowires is lowered, which is detrimental to the preparation of nanowires. Fig. 3(c) shows the extent to which the width of the nanowires obtained under different exposure dosage deviates from the theoretically predicted value (Deviation Factor, DF). The utilized equations are Eqs. (1) and (2):

$$\text{DF}_{100} = \left(\frac{|W_{max} - 100|}{100} + \frac{|W_{min} - 100|}{100}\right) \times 100\% \tag{1}$$

$$\text{DF}_{min} = \left(\frac{W_{max} - W_{min}}{W_{min}}\right) \times 100\% \tag{2}$$

$DF_{100}$ represents the total percentage of the maximum linewidth and the minimum linewidth offset the design values of 100 nm for a specific exposure dosage, reflecting both theoretical and practical consistency; $DF_{min}$ indicates the percentage of the difference between the maximum and the minimum linewidth at a specific exposure dosage, indicating the uniformity of the nanowires; $W_{max}$ and $W_{min}$ represent the maximum linewidth and the minimum linewidth, respectively. Fig. 3(b) and 3(c) show that the linewidths of the four pixels are around 100 nm at a dosage of 1400 μC/cm², the proximity effect is relatively weak, and the values of $DF_{100}$ and $DF_{min}$ are both minimal, indicating ideal exposure. The dosage was therefore set to 1400 μ/cm² in electron beam lithography. After the nanowire pattern formed, it was developed at room temperature using a ZX-238 type HSQ developer, an optimized development time of 2 min 30 s. Then, the NbN nanowires were obtained with a reactive ion etching apparatus RIE-10 (SAMCO), and $SF_6$ and $CHF_3$ were selected as etching gases with flow rates of 40 sccm and 10 sccm, respectively. The gas pressure was set to 4 Pa, and the etching time was 20 s.

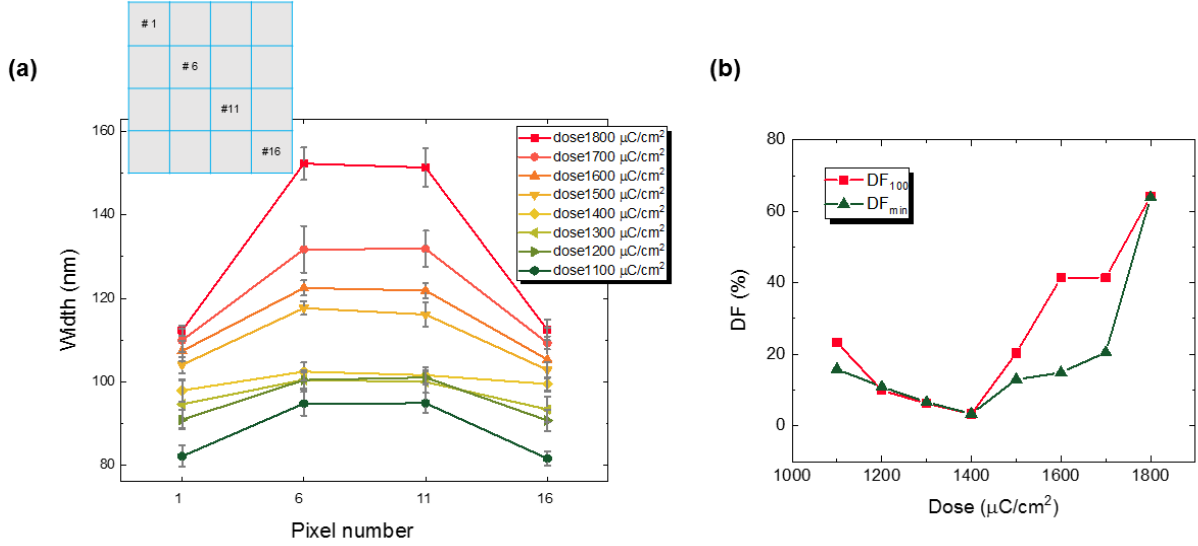

Fig. 3 (a) The inset shows the four pixels: pixel 1, pixel 6, pixel 11, and pixel 16, which were selected in the experiment. Nanowire linewidth and deviation for the four pixels change with dosage changes from 1100 μC/cm² to 1800 μC/cm². (b) The orange line shows the total deviation percentage of the nanowire maximum and minimum linewidths from the design values of 100 nm at different exposure dosage. The blue line shows the percentage of the difference between the minimum and maximum nanowire linewidths at different exposure dosage.

After preparation of the nanowire arrays, similar to the growth of the lower resonator, the upper $Si_3N_4$ optical cavity is grown by PECVD. Similar to the growth of the Au electrode, an Au mirror was grown via DC magnetron sputtering. The superconducting nanowires of all pixels are connected to the corresponding electrodes, and the electrodes are connected to the RF connector.

## 3. Results and analysis
**Measurement settings**

A dedicated package and opto-light platform were designed for the SNSPD, with one pin of 16 pixels end grounded, and the other end directed to the SMA port, respectively. To simultaneously transfer the bias current and voltage pulses, a lens was used to focus the signal from the multimode fiber on the active area of the SNSPD as shown in Fig. 4(a). Fig. 4(b) shows a schematic diagram of the signal processing of the SNSPD used in the experiment. $D_1, D_2,..., D_{15}$, and $D_{16}$ represent mutually independent pixels, $A_1, A_2,..., A_{15}$, and $A_{16}$ represent 16 identical low noise power amplifiers. The experimental light source adopts a picosecond pulse laser with an output light wavelength of $\lambda = 1064$ nm. The adjustable frequency of the laser is 1-100 MHz. Photons emitted by the laser are attenuated by a variable attenuator, then coupled into the fiber, and split into 16 channels via coupling lens. Polarization controllers were used to demonstrate that the efficiency of the detectors did not change with the light polarization, the SNSPD is polarization-insensitive detector. Once the optical signal enters the SNSPD, it is very quickly converted into an electrical signal that can be output by the bias circuit. The output signal is amplified by a low noise amplifier. The pulses of each pixel are finally collected, displayed via real-time oscilloscope, and analyzed with a digital counter. All optical components in the system are fiber-optic, and all electrical components are connected via coaxial cable.

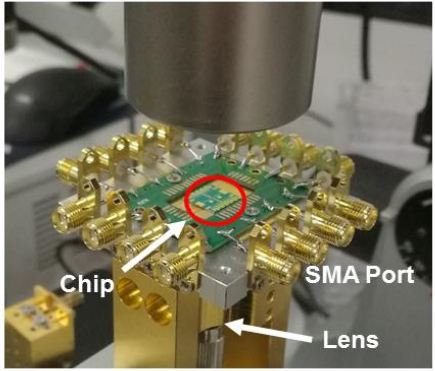
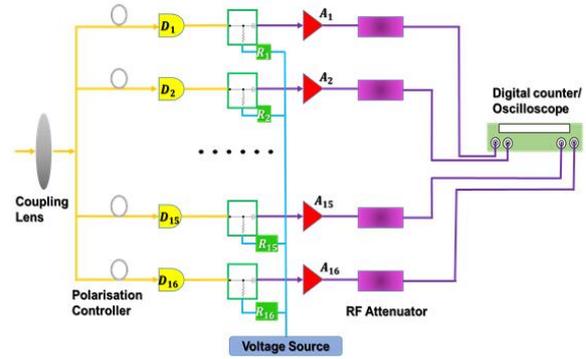

Fig. 4. (a) A dedicated package and opto-light platform were designed for the SNSPD, with one pin of 16 pixels end grounded, and the other end directed to the SMA port, respectively. (b) Schematic diagram of the signal processing for SNSPD.

**Consistency Analysis**

The average resistance value of the 16 ports of the array device measured at room temperature is 4.04 ± 0.15 MΩ with a relative standard deviation of 3.7%. The device resistance value was evenly distributed. The device was mounted on a compacted G-M cryocooler to cool down to 2.1 K. The mean value of the superconducting critical current $I_c$ of each pixel is 11.63 ± 0.8 µA with a relative standard deviation of 6.9%.

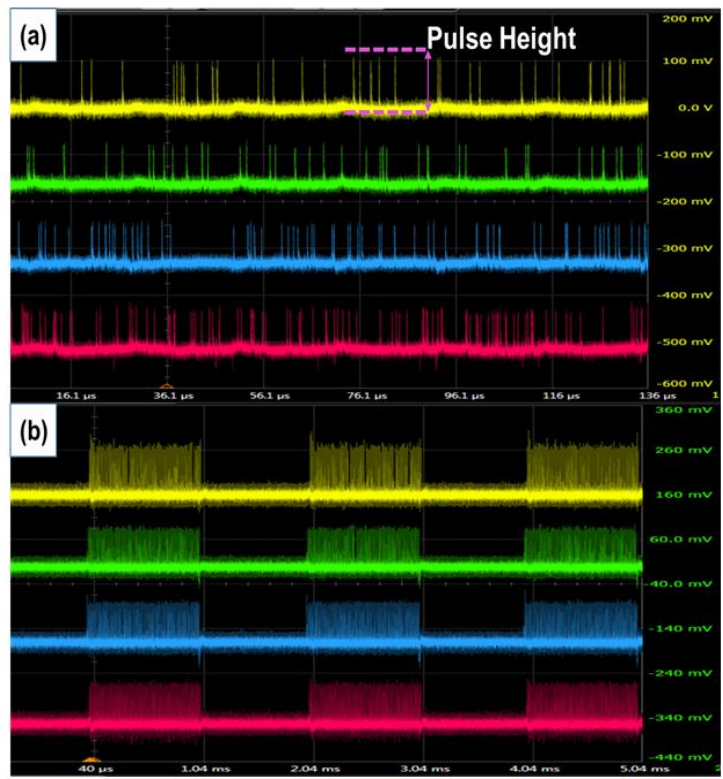

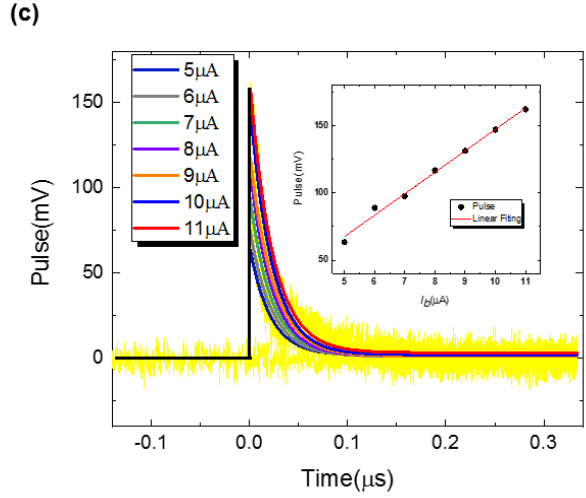

Fig. 5. Output signal generated by an arbitrarily selected four-channel. (a) Random distribution of the electrical pulse in run-free mode. (b) Electrical pulse signal in quasi-gated mode. (c) Voltage pulse exponential fitting curve for different bias currents. The inset shows linear Relationship between $V_p$ and $I_b$

A high-speed oscilloscope collects all electrical pulse signals generated by the nanowires after photon absorption when the system is in operation. Fig. 5 shows that the high-speed oscilloscope simultaneously outputs electrical pulse signals generated by the single photon response of the four arbitrarily selected channels. As shown in Fig. 5(a), the rising edge time of the generated electrical pulse signal is extremely short, and the recovery time is about 33 ns. Fig. 5(b) shows the random distribution of the electrical pulse signals recorded by the oscilloscope over an extended period of time. Fig. 5(c) shows that the oscilloscope further extends the recording time of the electrical pulse signal, using a gated method. Comparison shows that the amplitudes of the electrical pulse signals generated by the same pixel at different times are generally identical; furthermore, the amplitudes of the electrical pulse signals generated by different pixels are similar. This shows that the performance of all pixels is consistently and the system is stable. The bias current $I_b$ affects the amplitude $V_p$ of the output voltage pulse. Pulses of different amplitudes can be obtained by adjusting the magnitude of the bias current, and the $V_p$-$I_b$ relationship diagram is shown in Fig. 5(d). As shown in Fig. 5(f), the amplitude of the output pulse increases proportionally with the bias current. The relationship between the pulse amplitude $V_p$ and the bias current $I_b$ can be described with Eq (3):

$$V_p = (I_b - I_0) \times 50 \times G \tag{3}$$

Where 50 Ω represents the input end impedance of the amplifier, G represents the gain of the amplifier, and $I_0$ represents the minimum value of the current when the detector responds to photons. Fig. 5(f) shows the relationship between $V_p$ and $I_b$ when using the LNA650 amplifier with a fitting slope of 15.973. Considering the international system of units, the gain $G$ of the amplifier is calculated as 50.049 dB using Eq.(3), while the gain $G$ of the LNA650 in the product manual is 52 dB. The deviation is caused by several factors including the input impedance change of the amplifier, the error of the oscilloscope measurement and the sudden change because the small amount of data collected is within the tolerance range. Fig. 5(e) shows the voltage pulse exponential fitting curve for different bias currents. The fitting equation is Eq. (4):

$$y = y_0 + Ae^{-\frac{t}{\tau}} \tag{4}$$

The average reset time $\tau$ of the 6-time fitting is 20.25 ns, and the resistance $R$ is 50 Ω. According to Eq. (4) $\tau = Lk / R$, the kinetic inductance $L_k$ of the nanowire is calculated as ~1 µH.

**Efficiency and dark count**

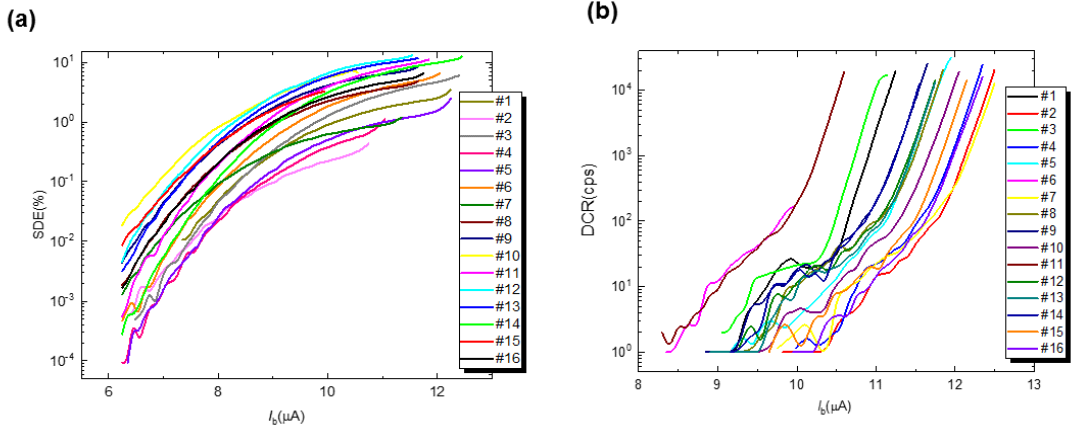

Fig. 6. The system detection efficiency (SDE) and dark count rate (DCR) were tested at the ultra-low temperature environment of 2.2 K, respectively. (a) Experimental results of the system detection efficiency. (b) Experimental results for the dark count rate.

The experiment used a pulsed laser to generate pulsed light of $\lambda$ = 1064 nm, and each pulse contained an average photon number of one. Since the experiment uses a multimode fiber with large spot size, the light intensity distribution can cover all pixels of the array device. The system detection efficiency (SDE) and dark count rate (DCR) are shown in Fig. 6 at an ultra-low temperature environment of 2.1 K. The SDE is defined as the ratio of the number of photons measured for each pixel to the total number of incident photons by considering the light coupling loss in this experiment. Fig. 6(a) shows that the SDE increases rapidly with increasing bias current $I_b$. Fig. 6(b) shows the three-dimensional columnar distribution of the SDE, showing the SDE of each pixel of the array device in different ranks. The total SDE of the system is the sum of the SDEs of each pixel, which reaches 46%. In addition, quantum efficiency (QE) was used to express the intrinsic efficiency of nanowires, which reflects the probability that a nanowire will generate and output a response electric pulse after absorption of one photon. Regardless of other secondary factors, QE is characterized in Eq. (5):

$$\text{QE} = \frac{SDE}{\alpha_0(1-\alpha_l)\eta_a\eta_c} \quad (5)$$

$SDE$ represents the total SDE which is saturated, $\alpha_0$, $\alpha_l$, $\eta_a$, and $\eta_c$ represent the transmittance of light through the silicon substrate, the loss rate of light in the multimode fiber, the absorption efficiency of photons by nanowires, and the coupling efficiency of multimode fibers, respectively. In the experiment, $SDE$ = 46%, $\alpha_0$ =71.1%, $\alpha_l$ = 8.8%, $\eta_a$ = 79%, $\eta_c$ = 95%, and QE = 94.5% was calculated. This indicates a high photoelectric conversion efficiency of the nanowires.

DCR refers to an electric pulse signal caused by the SNSPD system without a photon-reaction, resulting in an abnormal count that cannot be distinguished from the normal photon response. As shown in Fig. 6(c), when $I_b$ < 0.8$I_c$, the DCR of each pixel is below 10 cps, and the total SDE is below 24%. As the bias current $I_b$ remains close to the critical current $I_c$ of each pixel, the DCR of each pixel increases rapidly. Therefore, the working point of the device needs to be selected according to the actual application requirements, which cannot always be identical.

**Time jitter**

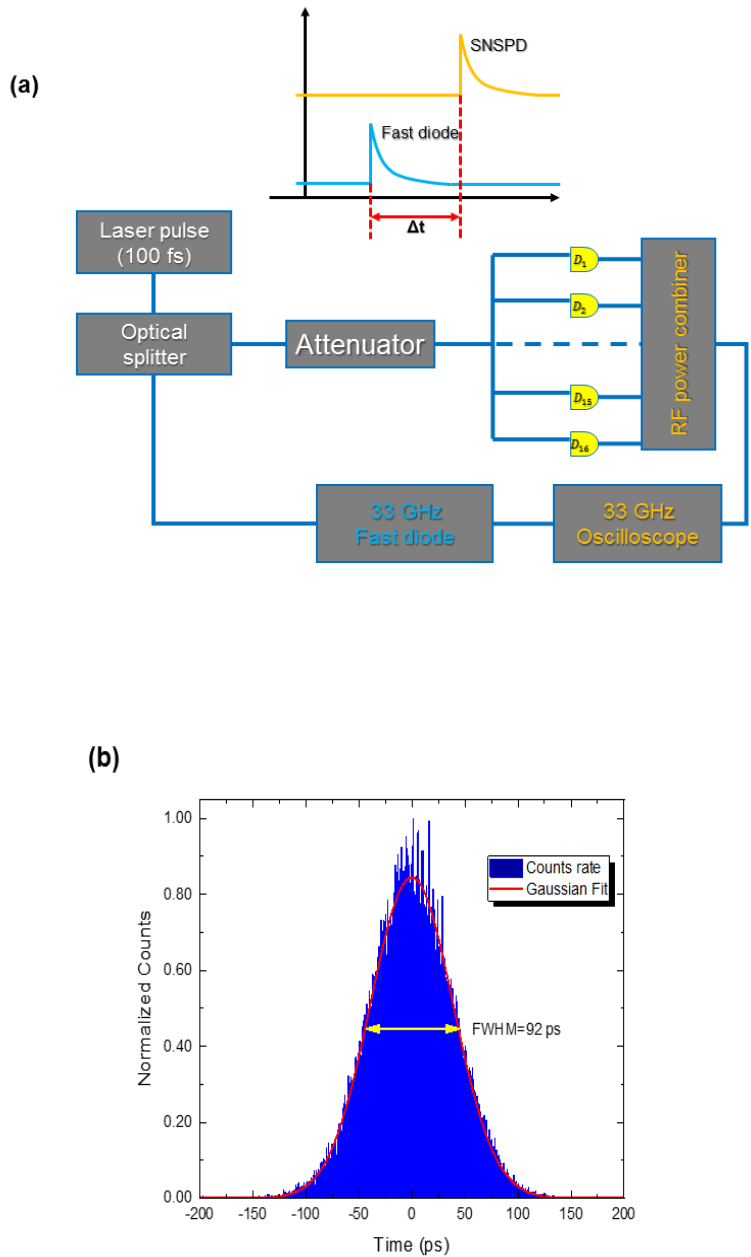

Fig. 7. (a) Time jitter measurement diagram. (b) The blue histogram shows the normalized count rate of photons, and the red curve show the Gaussian fit of the experimental data.

The time jitter is generally defined as the difference between the actual and the ideal arrival time of photons, or defined as the deviation of the interval between the rising edges of the electrical pulse signal generated by the single photon response and the theoretically predicted value when the pulse optical frequency remains constant.

In the experiment, the basic measurement system schematic diagram is shown in Fig. 7 (a). A mode-locked femtosecond laser with a pulse width below 50 fs and a repetition rate of 50 MHz was used. Jitter caused by the light pulse time distribution is negligible, since the light pulse is very thin. Furthermore, with a short fiber (a few meters long), the pulse broadening caused by dispersion can be ignored. The output light was split into two paths through the splitter, one can only be detected by a low-jitter fast diode, which gave a reference time signal on the high-speed oscilloscope. The other as the input signal of the SNSPD through the attenuator. Power synthesis[27] was used to superimpose response pulse signals from all channels, which are then output on the high-speed oscilloscope. A multimode fiber was used to measure the integrated time jitter at the wavelength of 1550 nm due to low $J_n$, low $J_0$, and low $J_{cd}$. The time jitter can be calculated by measuring the distribution of the delay (Δt) between the SNSPD response pulse and the reference pulse.

The time jitter is defined in Eq. (6):

$$J_{total} = \sqrt{J_n^2 + J_0^2 + J_{intrisinc}^2 + J_{cd}^2} \qquad (6)$$

$J_{total}$ represents the integrated time jitter; $J_n$ represents the time jitter generated by the circuit part; $J_0$ represents the inherent time jitter of the high-speed oscilloscope, $J_0 \leq 0.5$ ps; $J_{intrinsic}$ represents the intrinsic time jitter of the array device, which not only contains the intrinsic time jitter of each pixel but also an additional time jitter caused by mutual interference between pixels; $J_{cd}$ represents the time jitter generated by the light source and the optical path, which is related to the source bandwidth, the dispersion coefficient, and the transmission fiber. Generally, this ranges within sub-picoseconds. According to the linear small model perturbation theory and a previous publication[17], $J_n$ can be given by Eq. (7):

$$J_n = \frac{\sigma_n}{k} \times 2\sqrt{2ln2} \qquad (7)$$

Where $\sigma_n$ and $k$ represent the root mean square of the circuit noise and the slope of the rising edge of the response signal at the discrimination level, respectively. The magnitude of $\sigma_n$ is determined via the amplitude of the noise signal, $k$ is affected by the rising edge time of the pulse signal, the magnitude of the bias current, the gain of the amplifier, and the response bandwidth. As shown in Fig.7 (b), the blue histogram shows the normalized count rate of photons, and the red curve shows the Gaussian fit of the experimental data. $J_{total}$ is ~ 92 ps. According to statistical analysis, the magnitude of n is 5.077 mV, and the value of $k$ is 196 mV/ns, resulting in 61ps of $J_n$. $J_{intrinsic}$ < 69 ps according to Eq. (6).

## 4. Conclusion

In summary, a novel multi-pixel SNSPD was proposed and fabricated in this experiment. The developed 4×4 array consists of 16 compactly arranged pixels. The single pixel size is 20 μm × 20 μm, and 80 μm × 80 μm for the total active area. A multimode fiber with a diameter up to 300 μm is introduced into the optical path via beam compression technology, which greatly reduces the difficulty of optical coupling and the photon loss caused by this coupling process. The photon absorption efficiency does not change with the change of the polarization state of light. Loss caused by the unpredictable polarization state of the optical signal can be avoided when the device is used for space detection. The relationship between the exposure dosage and nanowire linewidth during electron beam exposure has been summarized in the presented experiments, and improved dosage conditions were obtained. The experimental results show that the system detection efficiency of the SNSPD reaches 46%, and a quantum efficiency of 94.5% at 1064 nm, indicating a high photoelectric conversion efficiency of the nanowires. The DCR of each pixel < 10 cps when $I_b$ (bias current) < $0.8I_c$ (critical current) and the device intrinsic time jitter < 69 ps. This work demonstrates the feasibility of large

SNSPDs array based on NbN, and paves a new way for developing large pixel photons camera.